\documentclass[12pt]{article}
\usepackage[super]{cite}
\usepackage{graphicx}
\usepackage{xspace}

\newcommand{\Npart}{\mbox{$N_{\rm part}$}\xspace}

\newcommand{\Nch}{\mbox{$N_{\rm ch}$}\xspace}
\newcommand{\Et}{\mbox{${\rm E}_T$}\xspace}

\newcommand{\sqs}{\mbox{$\sqrt{s}$}\xspace}
\newcommand{\sqsn}{\mbox{$\sqrt{s_{_{NN}}}$}\xspace}

\newcommand{\Nqp}{\mbox{$N_{qp}$}\xspace}
\def\lsim{\raise0.3ex\hbox{$<$\kern-0.75em\raise-1.1ex\hbox{$\sim$}}}
\def\gsim{\raise0.3ex\hbox{$>$\kern-0.75em\raise-1.1ex\hbox{$\sim$}}}

\def\mean#1{\left<#1\right>}
\def\Journal#1#2#3#4{{\it #1}{\bf #2}, #3 (#4)}
\def\IJMPA{{Int. J. Mod. Phys. A}\ }

\def\EPJC{{Eur. Phys. J. C\ }}

\def\MPLA{{Mod. Phys. Lett. A\ }}

\def\NPA{{Nucl. Phys. A\ }}
\def\NPB{{Nucl. Phys. B\ }}

\def\PL{Phys. Lett.\ }
\def\PRL{Phys. Rev. Lett.\ }
\def\PRD{{Phys. Rev. D\ }}
\def\PRC{{Phys. Rev. C\ }}

\def\ZPC{{Z. Phys. C\ }}
\def\ARNPS{{Ann. Rev. Nucl. Part. Sci.\ }} 
\def\RMP{Rev. Mod. Phys.\ }

\begin{document}

\markboth{M.~J.~Tannenbaum}
{Constituent quarks and systematic errors.}

\normalsize
\title{\large\bf Constituent quarks and systematic errors\\ in mid-rapidity charged multiplicity
$dN_{\rm ch}/d\eta$ distributions.}
\normalsize
\author{M.~J.~Tannenbaum
\thanks{Research supported by U.~S.~Department of Energy, DE-SC0012704.}
\\\it Physics Department, 510c, 
Brookhaven National Laboratory,\\
\it Upton, NY 11973-5000, USA\\
\it mjt@bnl.gov} 
\date{\normalsize Published December 29, 2017}
\maketitle

\begin{abstract}
Centrality definition in A$+$A collisions at colliders such as RHIC and LHC suffers from a correlated systematic uncertainty caused by the efficiency of detecting a p$+$p collision ($50\pm 5$\% for PHENIX at RHIC). In A$+$A collisions where centrality is measured by the number of nucleon collisions, $N_{\rm coll}$, or the number of nucleon participants, $N_{\rm part}$, or the number of constituent quark participants, $N_{\rm qp}$,  the error in the efficiency of the primary interaction trigger (Beam-Beam Counters) for a p$+$p collision leads to a correlated systematic uncertainty in $N_{\rm part}$, $N_{\rm coll}$ or $N_{\rm qp}$ which reduces binomially as the A$+$A collisions become more central. If this is not correctly accounted for in projections of A$+$A to p$+$p collisions, then mistaken conclusions can result. A recent example is presented in whether the mid-rapidity charged multiplicity per constituent quark participant $({dN_{\rm ch}/d\eta})/{N_{\rm qp}}$ in Au$+$Au at RHIC was the same as the value in p$+$p collisions.  
\end{abstract}

\section{\large From Nucleon Participants (Wounded Nucleons) to Constituent Quark Participants accounting for systematic errors}  
Measurements of charged particle multiplicity $N_{\rm ch}$ and transverse energy $E_T$ distributions in A$+$A collisions in the c.m. energy range  $10\leq\sqsn\leq 20$ GeV found the surprising result that the average charged particle multiplicity $\mean{N_{\rm ch}}$ in hadron+nucleus (h$+$A) collisions was not simply proportional to the number of collisions (absorption-mean-free-paths), $N_{\rm coll}$, but increased much more slowly, proportional to the number of nucleon participants, $N_{\rm part}$.\cite{Busza1975}$^,$ \cite{MJTIJMPA} This is known as the Wounded Nucleon Model (WNM)~\cite{WNM}.

\begin{figure}[!t] 
      \centering
 \includegraphics[width=0.9\textwidth]{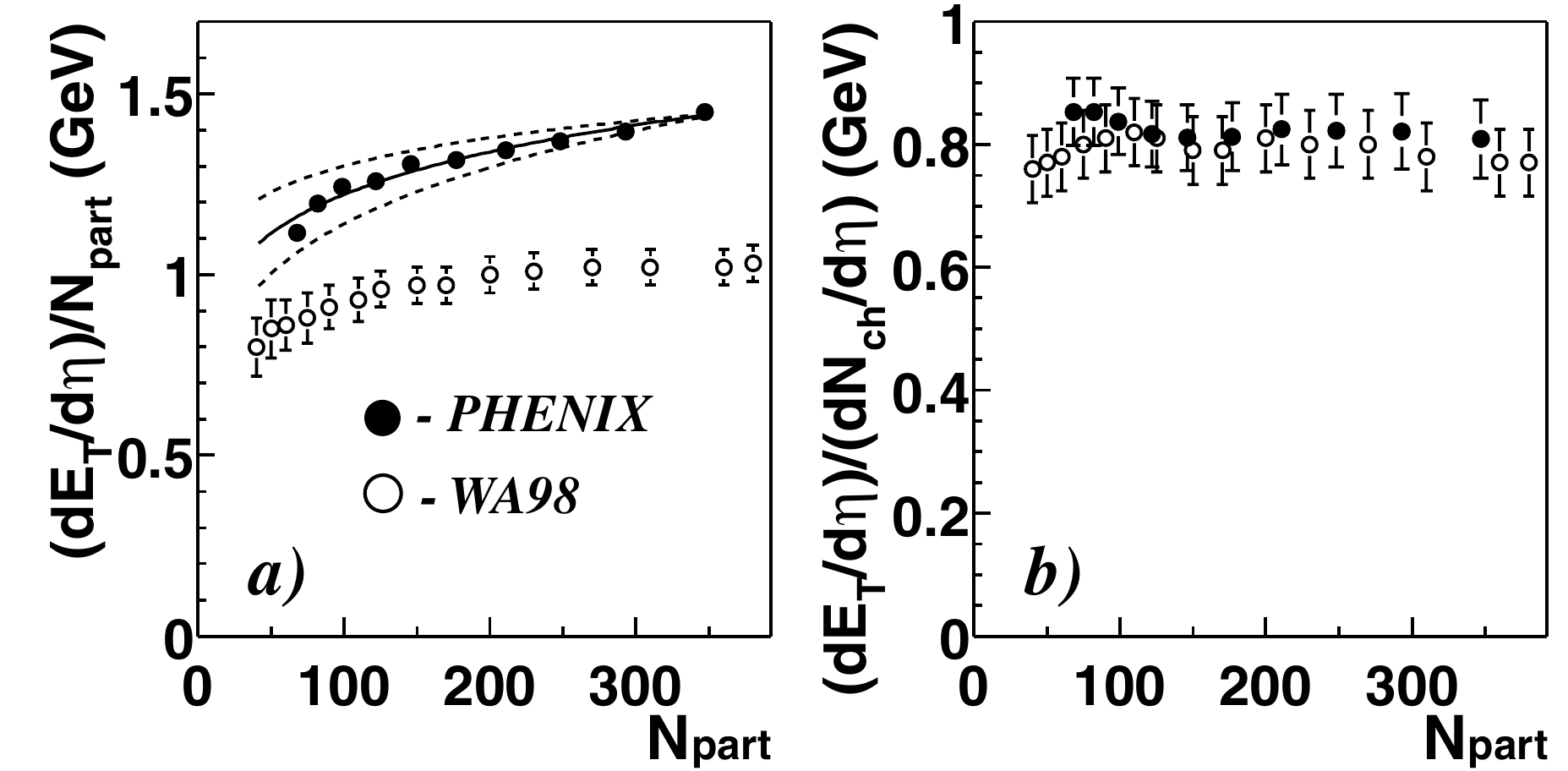}
      \caption[]{\footnotesize a)(left) PHENIX~\cite{ppg002} transverse energy density per participant $(d\Et/d\eta)/\Npart$ for Au$+$Au at \sqsn=130 GeV as a function of \Npart, compared to the data~\cite{WA98} of WA98 for Pb$+$Pb collisions at \sqsn=17.2 GeV. b) (right) ratio of $(d\Et/d\eta)/(d\Nch/d\eta)$.\label{fig:ppg002}} \vspace*{-0.5pc}
   \end{figure}

However, the first measurements of $d\Nch/d\eta$ (Ref.~4) and $d\Et/d\eta$ (Ref.~5) as a function of centrality at $\sqsn=130$ GeV at RHIC did not depend linearly on \Npart but had a non-linear increase of $d\Nch/d\eta$ and $d\Et/d\eta$ with increasing \Npart illustrated by the fact that $(d\Et/d\eta)/\Npart$ in Fig.~\ref{fig:ppg002}(a) is not constant but increases with \Npart. The comparison to the WA98 measurement~\cite{WA98} in Pb$+$Pb at \sqsn=17.2 GeV is also interesting because it is basically constant for \Npart$>200$, indicating that the WNM works at \sqsn=17.2 GeV in this region. 

The relevant point about the PHENIX $(d\Et/d\eta)/\Npart$ measurement for the present discussion (Fig.~\ref{fig:ppg002}a) is that the statistical errors are negligible. The dashed lines represent the $\pm 1 \sigma$ effect of the correlated systematic error. In PHENIX we call these Type B correlated systematic errors~\cite{ppg079}---all the points move together by the same fraction of the systematic error at each point. For instance, if all the points moved up to the dashed curve, which is $+1.0$ times the systematic error at each point, this will add only $1.0^2=+1.0$ to the value of $\chi^2$ of a fit with all 10 points  moved. 
The fit~\cite{ppg079} takes account of the statistical ($\sigma_i$) and correlated systematic ($\sigma_{b_i}$) errors for each data point with value $y_i$ : 
	\begin{equation}
{\chi}^2={\left[\sum_{i=1}^{n}
{{(y_i+\epsilon_b \sigma_{b_i} -y_i^{\rm fit})^2}  \over {{\tilde{\sigma}}_i^2}} \right]}+ {\epsilon_b^2} \qquad , 
\label{eqr:lstsq}
\end{equation}
where ${\tilde{\sigma}}_{i}$ scales the statistical error $\sigma_i$ by the shift in $y_i$ such that the
fractional error remains unchanged:
$\tilde{\sigma}_i=\sigma_i \left(1+\epsilon_b \sigma_{b_i}/{y_i}\right)$, where $\epsilon_b$ is to be fit.  

\subsection{The PHENIX2014 \Nqp model~\cite{PXPRC89}}
   \begin{figure}[!th] 
      \centering
 \includegraphics[width=0.9\textwidth]{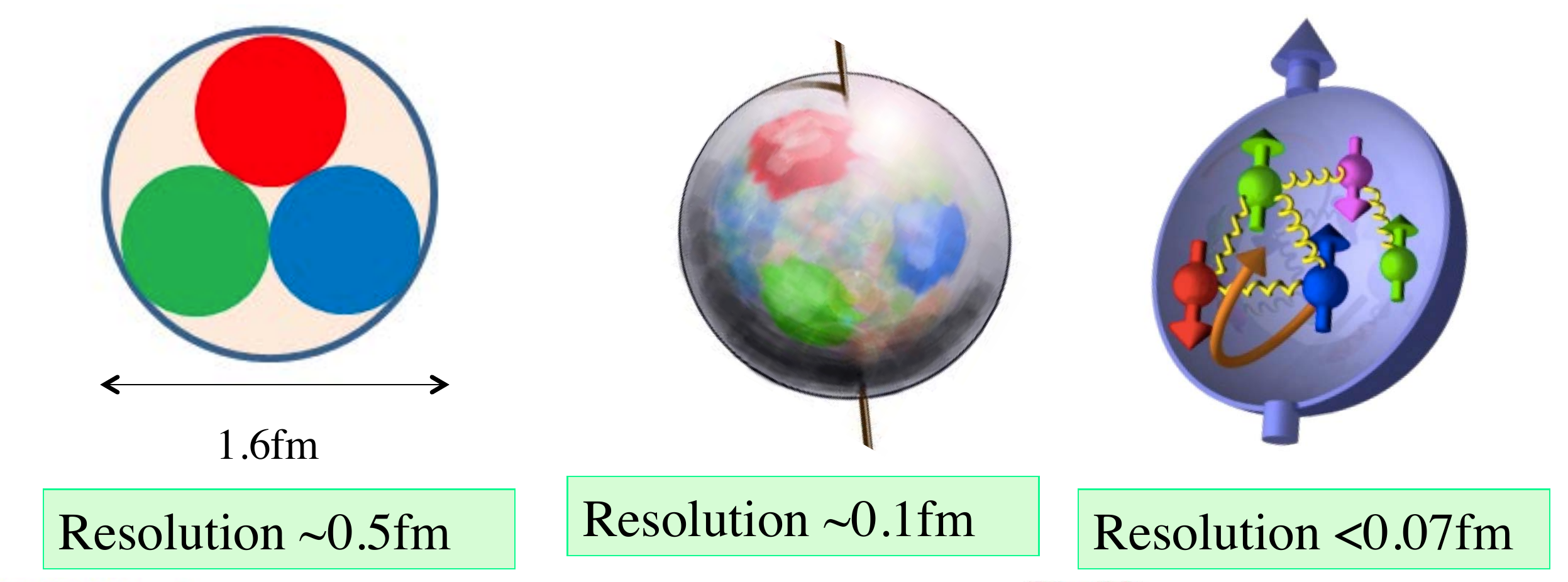}
      \caption[]{\footnotesize Sketch of constituent quarks in a proton (diameter 1.6 fm) with improving resolution: (left) $\approx 0.5$ fm, (center) $\approx 0.1$ fm, (right) $< 0.07$ fm, massless quarks, sea quarks and gluons clearly resolved\label{fig:resolution}}\vspace*{-0.4pc} 
   \end{figure}
	The massive constituent quarks~\cite{MGMPL8,MuellerNPA527,MorpurgoRNC33}, which form mesons and nucleons (e.g. a proton=$uud$), are relevant for static properties and soft physics such as multiplicity and \Et distributions composed predominantly of particles with $p_T\lsim1.4$ GeV/c in p$+$p collisions. Constituent quarks are complex objects or quasiparticles~\cite{ShuryakNPB203} made of the massless partons (valence quarks, gluons and sea quarks) of DIS~\cite{DIS2} such that the valence quarks acquire masses $\approx 1/3$ the nucleon mass with radii $\approx 0.3$ fm when bound in the nucleon. With  finer  resolution (Fig.~\ref{fig:resolution}) one can see inside the bag to resolve the massless partons which can scatter at large angles according to QCD. At RHIC, hard-scattering starts to be visible as a power law above soft (exponential) particle production at mid-rapidity only for $p_T>$ 1.4 GeV/c~\cite{PXpi0PRD}, where $Q^2=2p_T^2=4$ (GeV/c)$^2$ which corresponds to a distance scale (resolution) $<0.1$ fm. 

The first 3 calculations which showed that \Nch was linearly proportional to~ \cite{EreminVoloshinPRC67,NouicerEPJC49,DeBhattPRC71} \Nqp only studied Au$+$Au collisions and simply generated three times the number of nucleons according to the Au radial disribution, Eq.~\ref{eq:WS}, 
\begin{equation}
\frac{d^3 {\cal P} } {d^3 r} =\frac{d^3 {\cal P}} {r^2 dr \sin\!\theta d\theta\,  d\phi}=\rho_{N}(r)=\frac{\rho_0}{1 + \exp (\frac{r-c}{a_0})} \label{eq:WS}
\end{equation}
with $c=\{1.18 A^{1/3} - 0.48\}$ fm and the diffusivity $a_0=0.545$ fm for Au,  
called them constituent quarks and let them interact with the conventional constituent $q+q$ cross section $\sigma^{\rm inel}_{q+q}=\sigma^{\rm inel}_{N+N}/9$, e.g $\sigma^{\rm inel}_{q+q}$=41mb/9=4.56 mb at \sqsn=130 GeV~\cite{EreminVoloshinPRC67}. 
\vspace*{0.2pc}

\begin{figure}[!t] 
      \centering
 \includegraphics[width=0.48\textwidth]{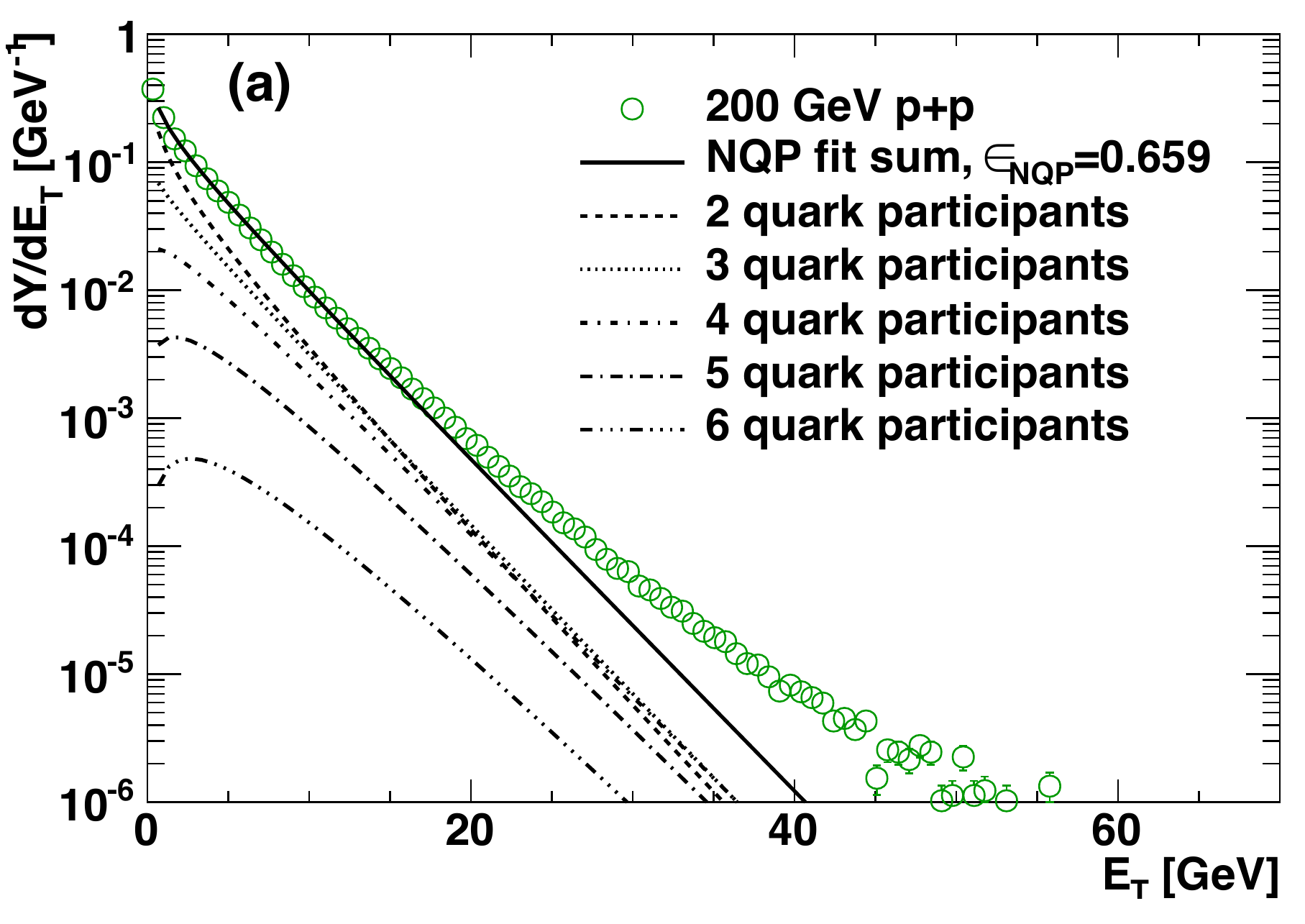}
      \includegraphics[width=0.48\textwidth]{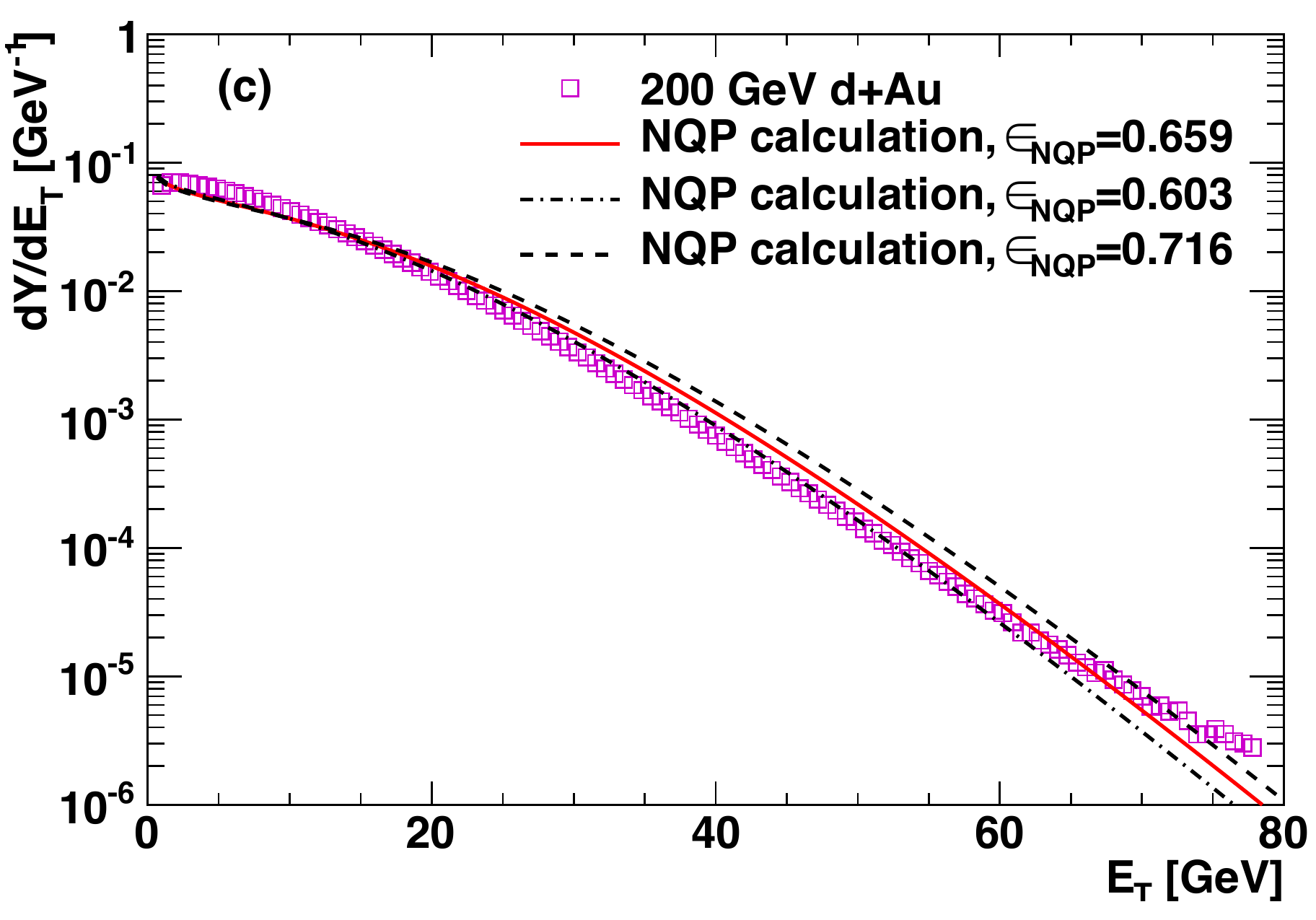}
      \includegraphics[width=0.70\textwidth]{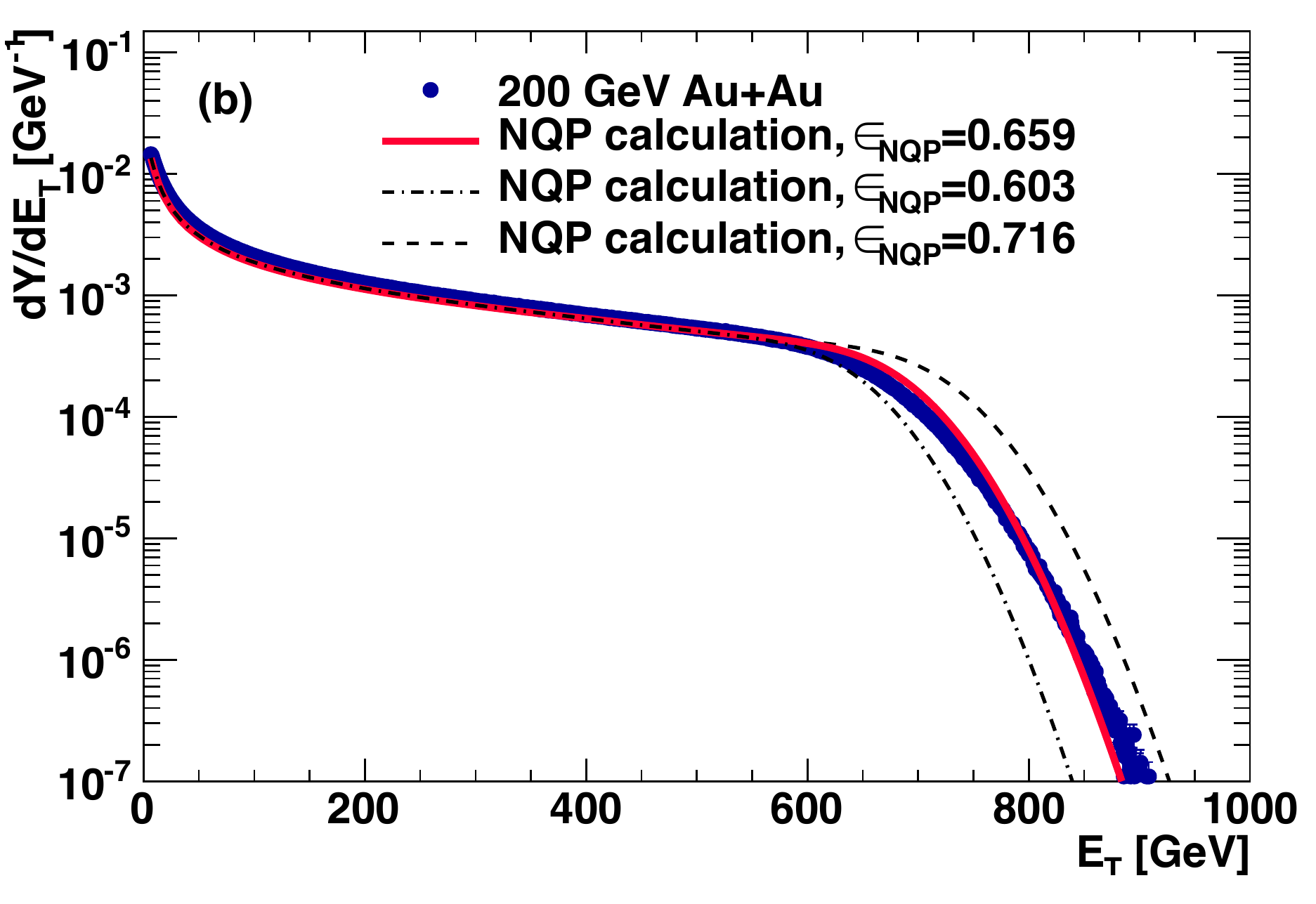}
      \vspace*{-0.5pc} 
      \caption[]{\footnotesize PHENIX2014~\cite{PXPRC89} method for $\Et\equiv d\Et/d\eta|_{y=0}$ distributions at $\sqsn=200$ GeV:
      a) Deconvolution fit to the $p$+$p$ \Et distribution for $\Et<13.3$ GeV for $\epsilon_{\rm NQP}=1-p_{0_{\rm NQP}}=0.659$ calculated in the Number of Quark Participants or \Nqp model. Lines represent the properly weighted individual \Et distributions for the underlying 2,3,4,5,6 constituent quark participants plus the sum.  b) Au+Au \Et distribution compared to the \Nqp calculations using the central $1-p_0=0.647$ and $\pm 1\sigma$ variations of $1-p_0=0.582,0.712$ for the probability $p_0$ of getting zero \Et on a $p$+$p$ collision with resulting quark participant efficiencies $\epsilon_{\rm NQP}=0.659,0.603,0.716$, respectively. (c) d+Au \Et distribution compared to the \Nqp calculations as in (b).        \label{fig:PXppg100}} \vspace*{-0.5pc}
   \end{figure}

The PHENIX2014 method~\cite{PXPRC89} was different from these \Nqp calculations in that it used the \Et distribution measured in p$+$p collisions to derive the \Et distribution of a constituent quark to use as the basis of the calculations of the $d+$Au and Au$+$Au distributions. The PHENIX2014 calculation~\cite{PXPRC89} is a Monte Carlo which starts by generating the positions of the nucleons in each nucleus of an A$+$B collision, or simply the two nucleons in a p$+$p collision, by the standard method. Then the spatial positions of the three quarks are  generated around the position of each nucleon using the proton charge distribution corresponding to the Fourier transform of the form factor of the proton~\cite{HofstadterRMP28,HofstadterRMP30}:
\begin{equation}
   \rho^{\rm proton}(r) = \rho^{\rm proton}_{0} \times \exp(-ar),
   \label{eq:Hofstadterdipole}
\end{equation}
where $a = \sqrt{12}/r_{m} = 4.27$ fm$^{-1}$ and 
$r_{m}=0.81$ fm is  
the r.m.s radius of the proton weighted according to charge~\cite{HofstadterRMP28} 
\begin{equation}
r_{m}=\int_0^\infty r^2 \times 4\pi r^2 \rho^{\rm proton}(r) dr \qquad .
\label{eq:rmsintegral}
\end{equation}
The corresponding proton form factor is the Hofstadter dipole fit~\cite{HandMillerWilson} now known as the standard dipole~\cite{BernauerMainzPRC90}:
\begin{equation}
G_E(Q^2)=G_M(Q^2)/\mu=\frac{1}{(1+\frac{Q^2}{0.71 {\rm GeV}^2})^2} \label{eq:HofstadterdipoleFF}
\end{equation}
where $G_E$ and $G_M$ are the electric and magnetic form factors of the proton, $\mu$ is its magnetic moment and $Q^2$ is the four-momentum-transfer-squared of the scattering.
The inelastic $q+q$ cross section $\sigma^{\rm inel}_{q+q}=9.36$ mb at \sqsn=200 GeV was derived from the p$+$p \Nqp Glauber calculation by requiring the  calculated p$+$p inelastic cross section to reproduce the measured $\sigma^{\rm inel}_{N+N}=42$ mb cross section,  and then used for the Au$+$Au and $d+$Au  calculations (Fig.~\ref{fig:PXppg100})~\cite{PXPRC89}. 

\begin{figure}[!t] 
      \centering
\raisebox{0.0pc}{\includegraphics[width=0.54\linewidth]{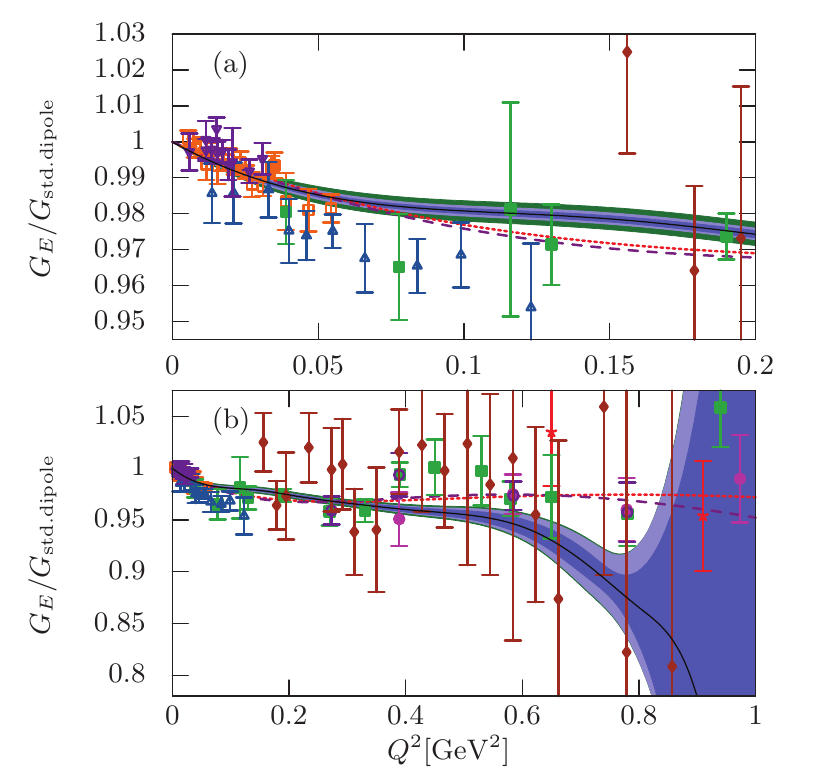}}
\raisebox{0.0pc}{\includegraphics[width=0.42\linewidth]{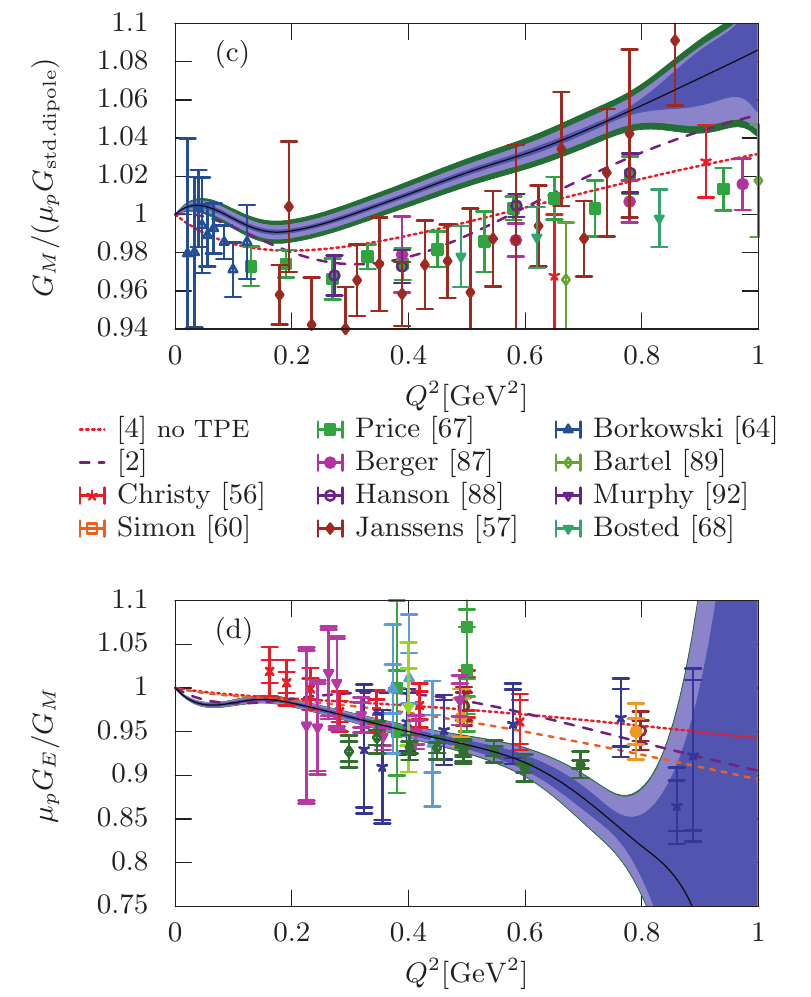}}\vspace*{-0.5pc} 
      \caption[]{\footnotesize The form factors $G_E$ and $G_M$, normalized to the standard dipole, and $G_E/G_M$, compared to fits, with the dark region being the best fit to the new Mainz data~\cite{BernauerMainzPRC90}. }\vspace*{-0.5pc}
      \label{fig:Mainz}
   \end{figure}

Sometimes people ask why we use Hofstadter's 60 year old measurements when there are more modern measurements which give a different proton r.m.s charge radius~\cite{BernauerMainzPRC90}, which is not computed from Eq.~\ref{eq:rmsintegral} but merely from the slope of the form factor at $Q^2=0$. The answer is given in Fig.~\ref{fig:Mainz} which shows how all the measurements of $G_E(Q^2)$ and $G_M(Q^2)$ for $Q^2\leq  1$ GeV$^2$ agree with the ``standard dipole'' (Eq.~\ref{eq:HofstadterdipoleFF}) within a few percent, and in all cases in Fig.~\ref{fig:Mainz} agree as well if not better than the Mainz fit. 
\subsection{Improved method of generating constituent quarks}
A few months after PHENIX2014 was published,~\cite{PXPRC89} it was pointed out to us that our method did not preserve the radial charge distribution (Eq.~\ref{eq:Hofstadterdipole}) about the c.m. of the three generated quarks. This statement is correct; so a few of us got together and found three new methods that preserve both the original proton c.m. and the correct charge distribution about this c.m.~\cite{CQ2016}. I discuss two of them here along with \Nqp calculations using the PHENIX2014 data.~\cite{PXPRC89}
\subsubsection{Planar Polygon}
Generate one quark at $(r,0,0)$ with $r$ drawn from $r^2 e^{-4.27r}$. Then instead of generating $\cos\theta$ and $\phi$ at random and repeating for the two other quarks as was done by PHENIX2014~\cite{PXPRC89}, imagine that this quark lies on a ring of radius $r$ from the origin and  place the two other quarks on the ring at angles spaced by $2\pi/3$ radians.  Then randomize the orientation of the 3-quark ring spherically symmetric about the origin.  This guarantees that the radial density distribution is correct about the origin and the center of mass of the three quarks is at the origin but leaves the three-quark-triplet on each trial forming an equilateral triangle on the plane of the ring, which passes through the origin.   
\subsubsection{Empirical radial distribution, recentered}
\begin{figure}[!t] 
      \centering
\raisebox{0.0pc}{\includegraphics[width=0.49\linewidth]{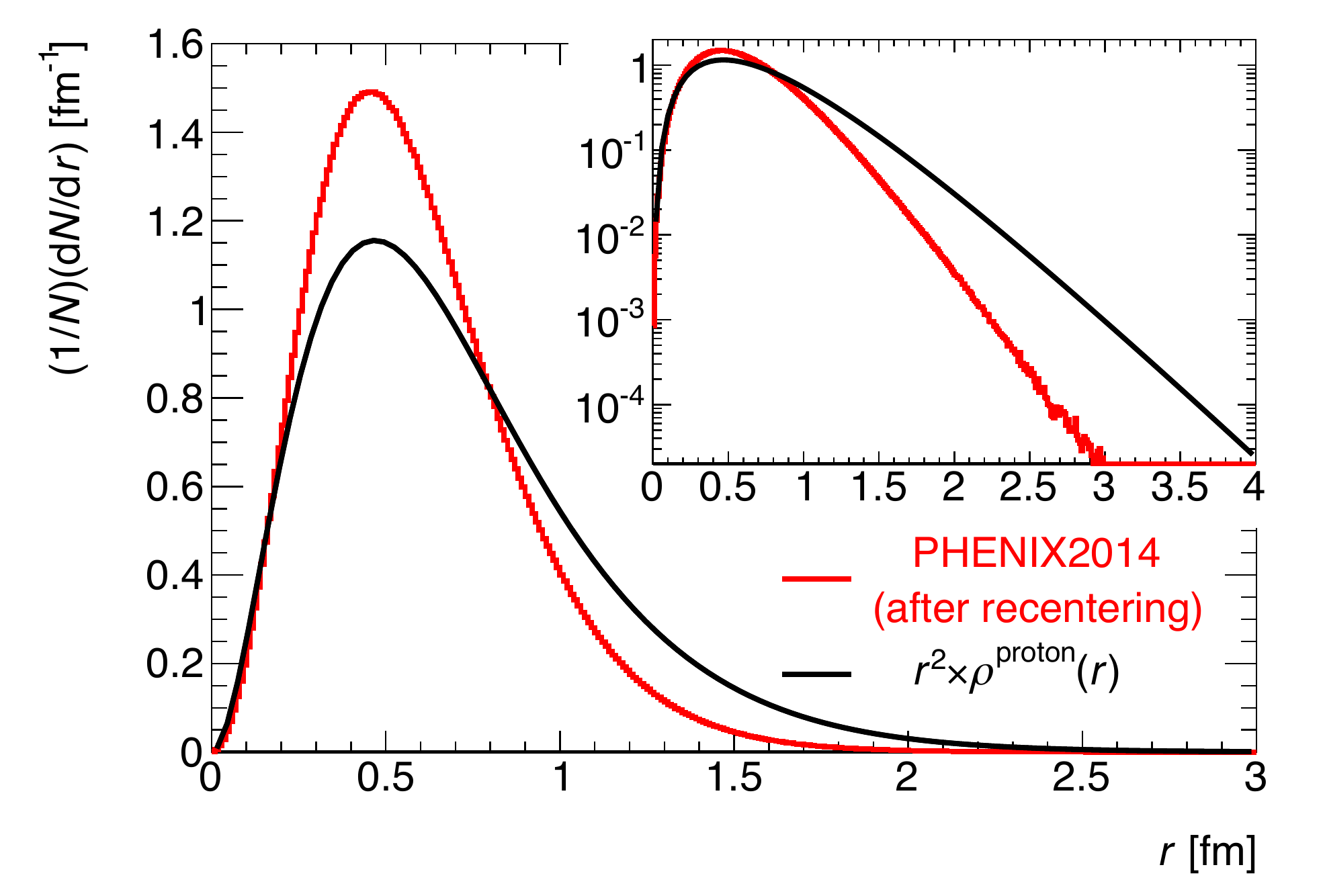}}
\raisebox{0.0pc}{\includegraphics[width=0.43\linewidth]{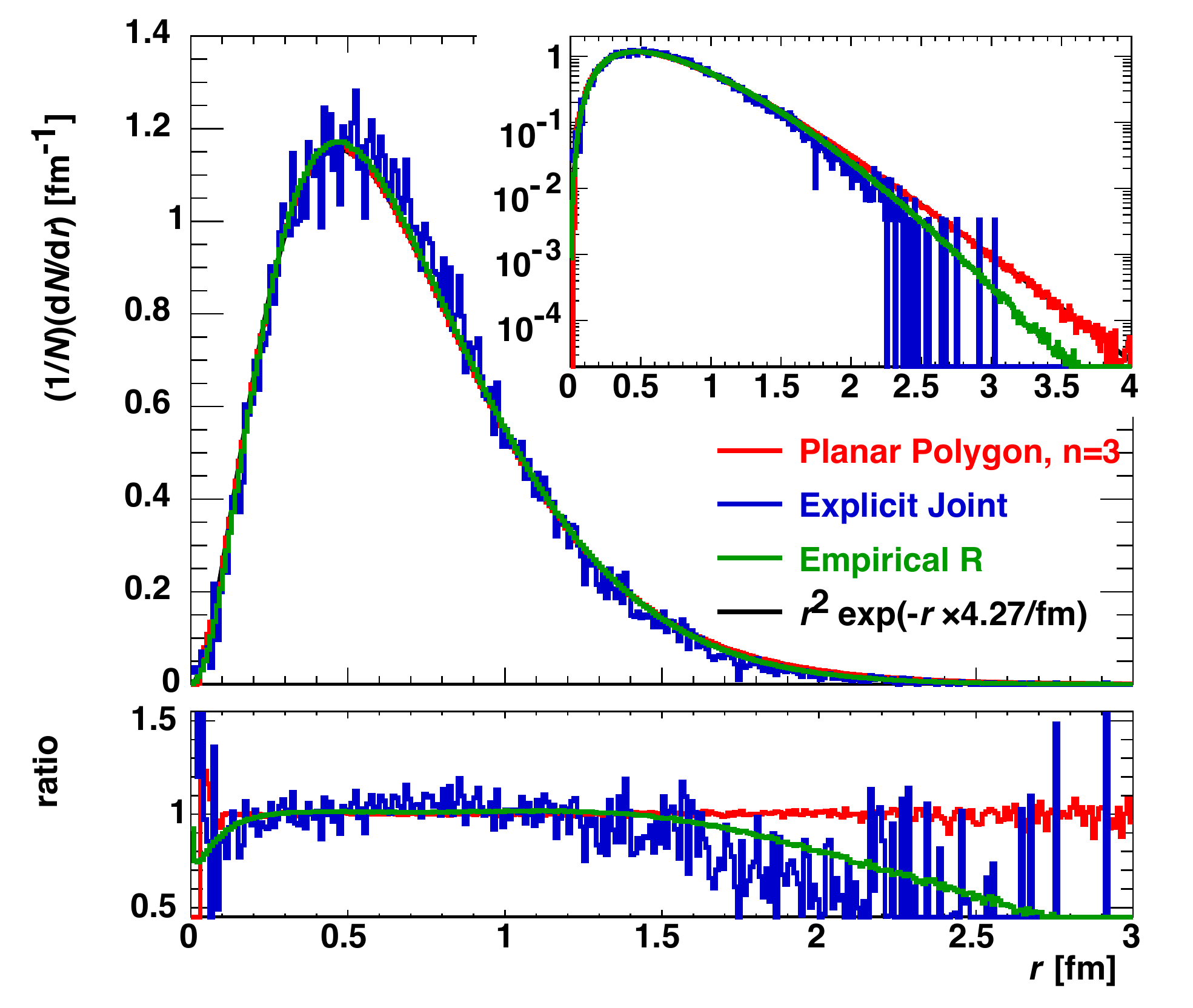}}\vspace*{-1.0pc} 
      \caption[]{\footnotesize a) (left) Radial distribution $d{\cal P}/dr$ about the c.m. of the generated quark-triplets  as a function of $r$ [fm] for the PHENIX2014 method~\cite{PXPRC89} compared to $r^2\rho^{\rm proton}(r)$ from Eq.~\ref{eq:Hofstadterdipole} with semi-log plot as inset. b) (right) same for the 3 new methods and the ratios as indicated~\cite{CQ2016}. }
      \label{fig:radial}\vspace*{1.0pc}
   \end{figure}

The three constituent quark positions are drawn independently from an auxiliary function $f(r)$:
\begin{equation}
   f(r)= r^2 \rho^{\rm proton}(r)\; (1.21466-1.888r+2.03r^{2})\;(1+1.0/r - 0.03/r^{2})\; (1+0.15r) \ .
\label{eq:empirical}
\end{equation}
Then the center of mass of the generated three-quark system is re-centered to the original nucleon position. This function was derived through an iterative, empirical approach. For a given test function $f^{\rm test}(r)$, the resulting radial distribution $\rho^{\rm test}(r)$ was compared to the desired distribution $\rho^{\rm proton}(r)$ in Eq.~\ref{eq:Hofstadterdipole}. The ratio of $\rho^{\rm test}(r) / \rho^{\rm proton}(r)$ was parameterized with a polynomial function of $r$ or $1/r$, and the test function was updated by multiplying it with this parametrization of the ratio. Then, the procedure was repeated with the updated test function $f^{\rm test}(r)$ used to generate an updated $\rho^{\rm test}(r)$ 
until the ratio $\rho^{\rm test}(r) / \rho^{\rm proton}(r)$ was sufficiently close to unity over a wide range of $r$ values. Figure~\ref{fig:radial} shows~\cite{CQ2016} the generated radial distributions compared to $r^2\rho^{\rm proton}(r)$ from Eq.\ref{eq:Hofstadterdipole}.
\subsection{New \Nqp results using PHENIX2014 data}   
\begin{figure}[hbt] 
      \centering
 \includegraphics[width=0.48\textwidth]{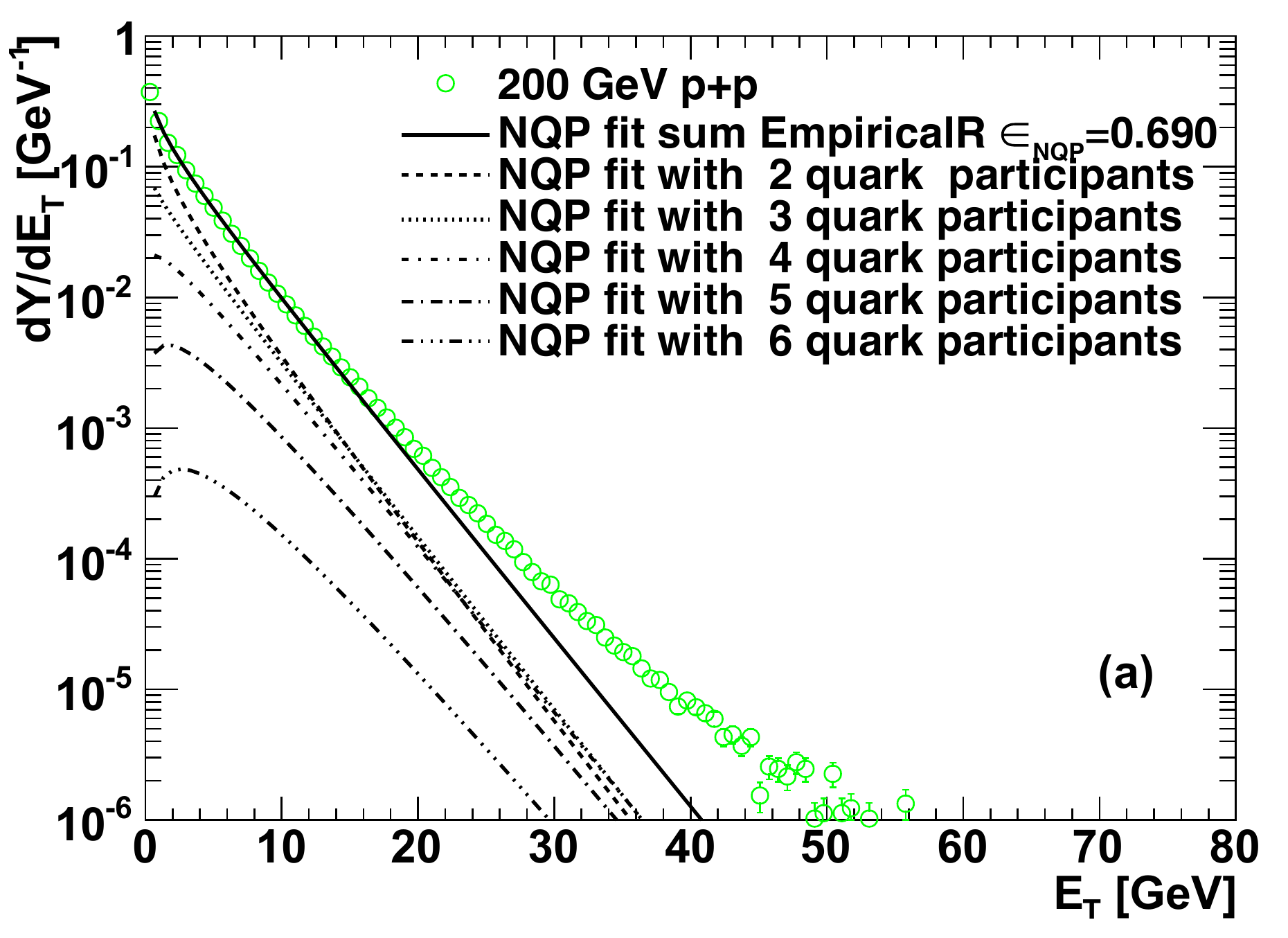}
      \includegraphics[width=0.48\textwidth]{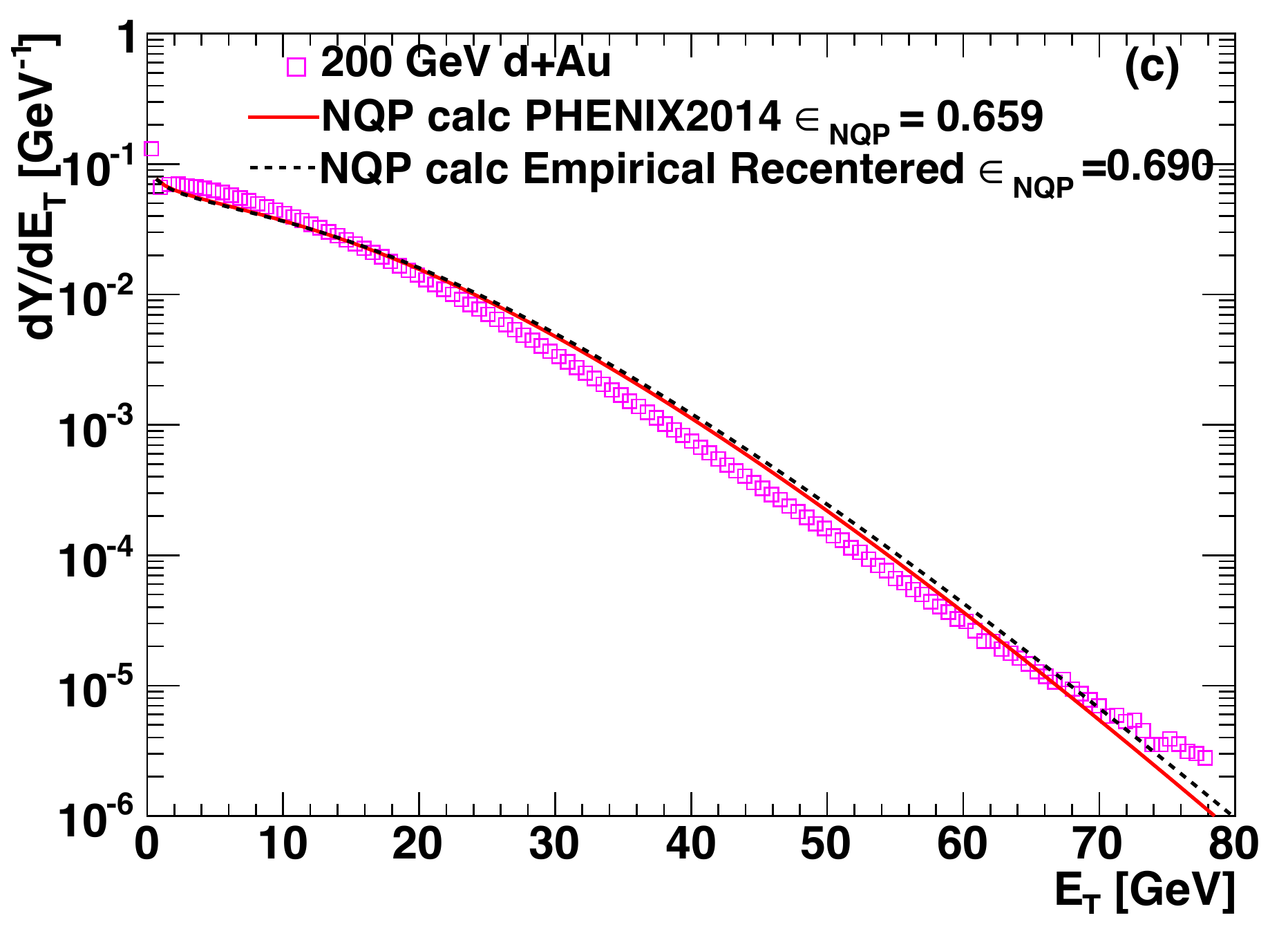}
      \includegraphics[width=0.70\textwidth]{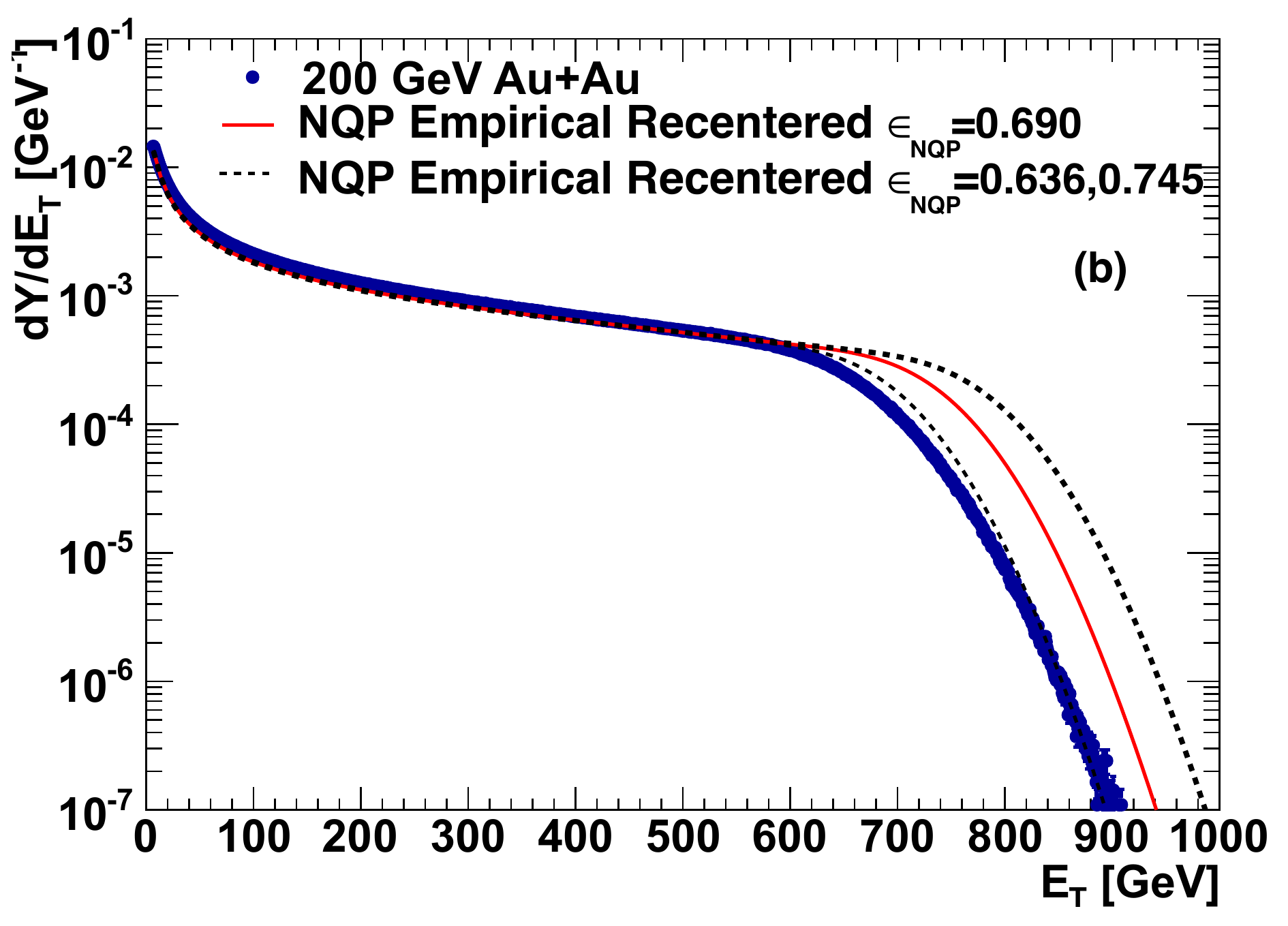}
      \vspace*{-0.5pc} 
      \caption[]{\footnotesize New \Nqp results~\cite{CQ2016} for $\Et\equiv d\Et/d\eta|_{y=0}$ distributions at $\sqsn=200$ GeV using and compared to PHENIX2014\cite{PXPRC89} data. a) p$+$p, b) Au$+$Au, c) d$+$Au.}
\label{fig:newNQP}
   \end{figure}

From Fig.~\ref{fig:radial}b, the Planar Polygon method is identical to Eq.~\ref{eq:Hofstadterdipole} but has all three quarks at the same radius from the c.m. of the proton, which can be tested with more information about constituent quark correlations in a nucleon. The Empirical Recentered method follows $r^2\rho^{\rm proton}(r)$ well out to nearly $r=2$ fm, $Q^2=0.25$ fm$^{-2}=0.01$ GeV$^2$ (compare Fig.~\ref{fig:Mainz}a,b), and is now adopted as the standard~\cite{Loizides2016}. The results of the \Nqp calculations with the Empirical Recentered method~\cite{CQ2016} for the PHENIX2014 data (Fig.~\ref{fig:newNQP}), are in excellent agreement  with the d$+$Au data and agree with the Au$+$Au measurement to within $1 \sigma$ of the calculation (7\% higher in \Et).     
The PHENIX2014 calculation (Fig.~\ref{fig:PXppg100}b) is only $1.2\sigma$ in \Et below the new calculation so that the PHENIX2014 \Nqp results and conclusions~\cite{PXPRC89} are consistent with the new standard method~\cite{CQ2016}. 
\section{\large Constituent quark participant scaling vs. centrality for Multiplicity and \Et distributions}
\begin{figure}[hbt] 
   \label{fig:p12r}
      \begin{center}
 \includegraphics[width=0.9\textwidth]{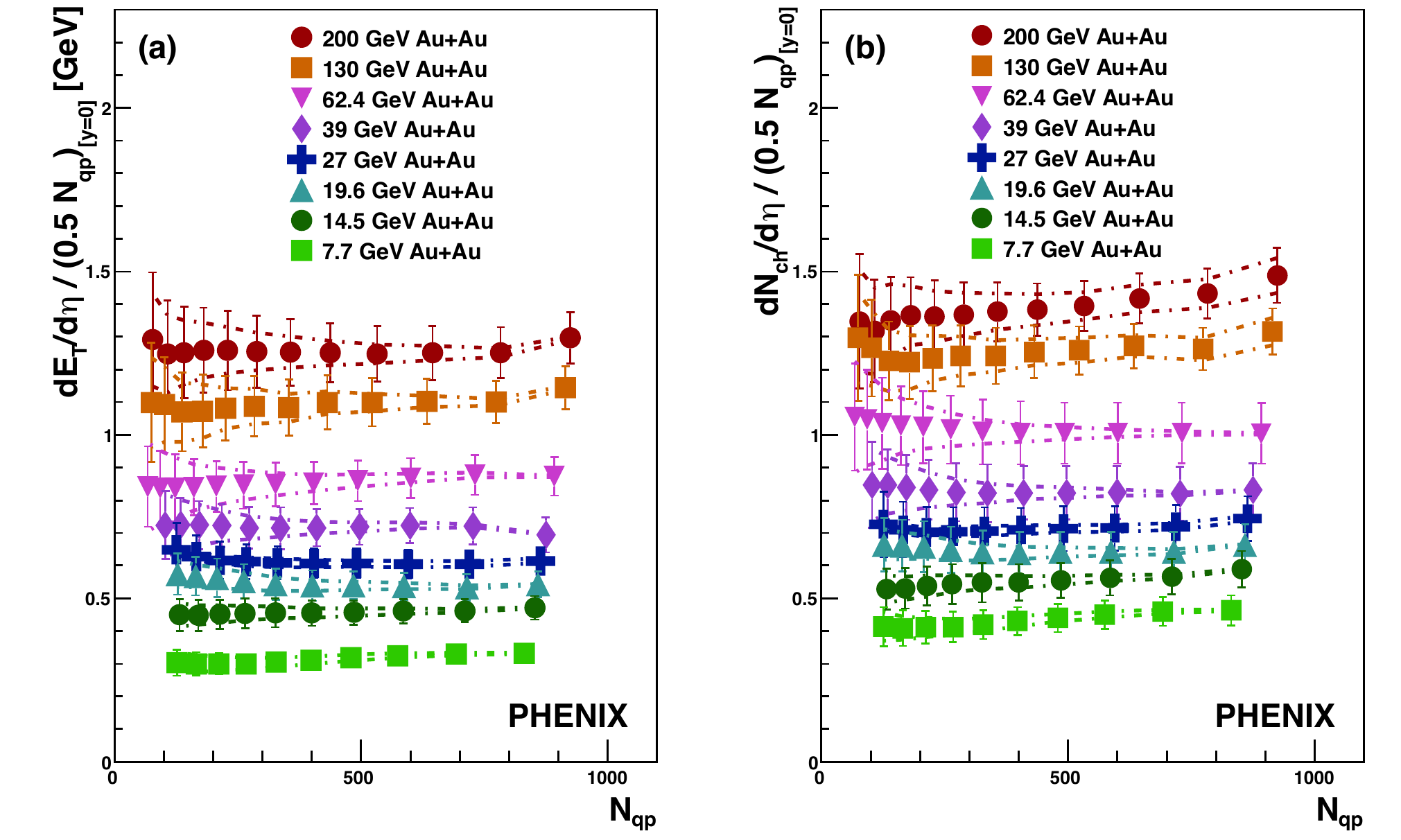}
      \vspace*{-0.5pc} 
      \caption[]{\footnotesize (a) $(d\Et/d\eta)/(0.5\Nqp)$ and  (b) $(d\Nch/d\eta)/(0.5\Nqp)$ at midrapidity as a function of \Nqp for Au$+$Au collisions at \sqsn=200, 130, 62.4, 39, 27, 19.6, 14.5, and 7.7 GeV. {\it The lines bounding the points represent the trigger efficiency uncertainty within which the points can be tilted}. The error bars represent the remaining total statistical and systematic uncertainty.~\cite{ppg174}} \label{fig:ppg174}
      \end{center}
   \end{figure}

In Fig.~\ref{fig:ppg174}~\cite{ppg174}, the linear dependence of $d\Et/d\eta$ and $d\Nch/d\eta$ on the number of constituent quark participants \Nqp is demonstrated by the constant values of $(d\Et/d\eta)/(0.5\Nqp)$ and $(d\Nch/d\eta)/(0.5\Nqp)$ with centrality represented as \Nqp.
The \Nqp calculations are made with the now standard Empirical Recentered method~\cite{CQ2016}. The relevant issue for the present discussion is to notice the $\pm1\sigma$ correlated systematic errors indicated by the dashes. For instance, in Fig.~\ref{fig:ppg174}b all the data points for 200 GeV Au$+$Au can be moved up by $+1\sigma$ of the correlated systematic error to the dashed line at the cost of only $1.0^2=+1.0$ to the value of $\chi^2$ of a fit with all 10 points moved. In detail this means that the lowest $(d\Nch/d\eta)/(0.5\Nqp)=1.35\pm0.20$ data point at \Nqp=78 can be moved up to the dashed line as $1.50\pm0.22$ and all the other data points will move up to the dashed line.  
\subsection{Disagreement from another \Nqp calculation?} 
\label{sec:Bozek}
\begin{figure}[hbt] 
      \centering
\includegraphics[width=0.32\textwidth]{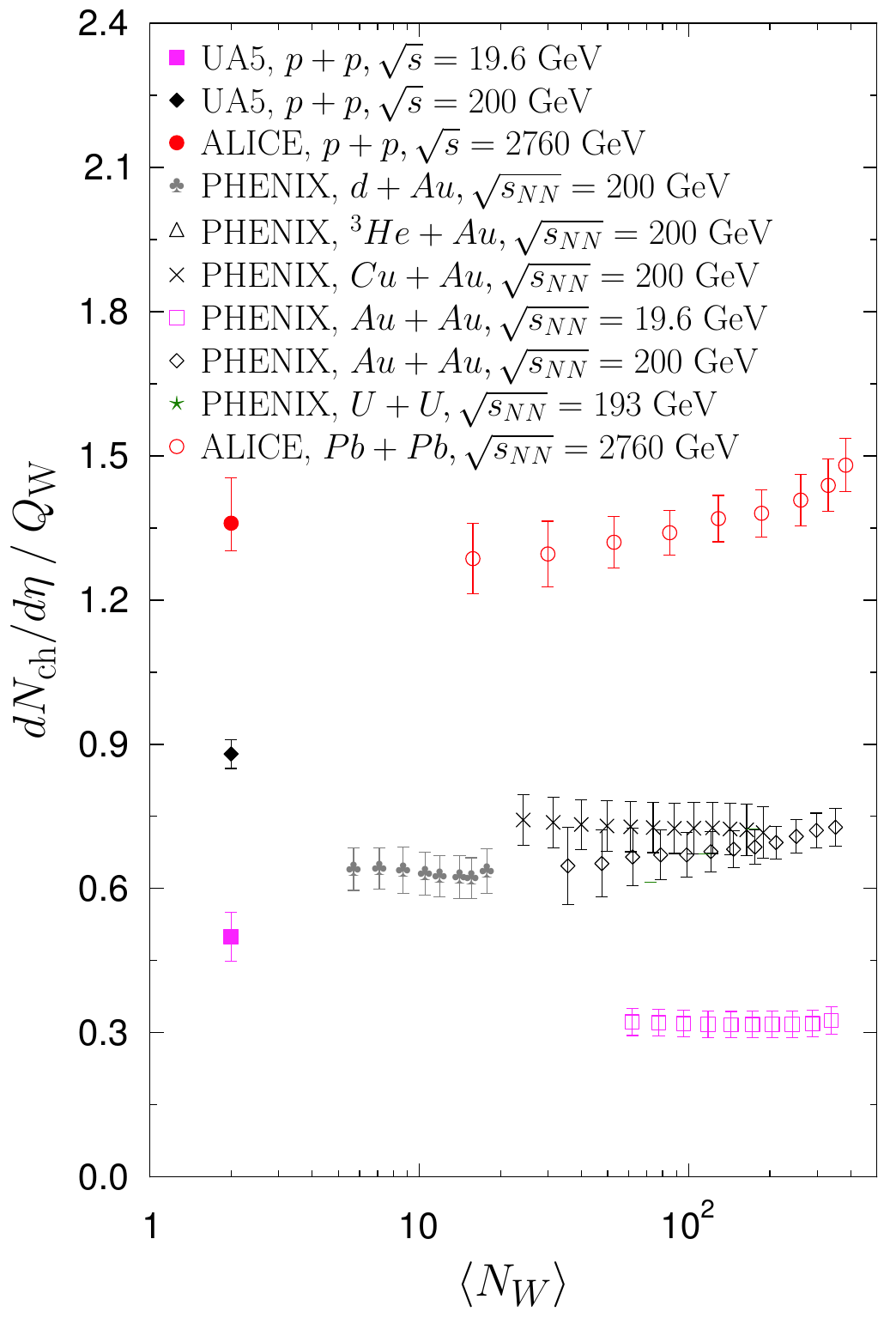}
\includegraphics[width=0.32\textwidth]{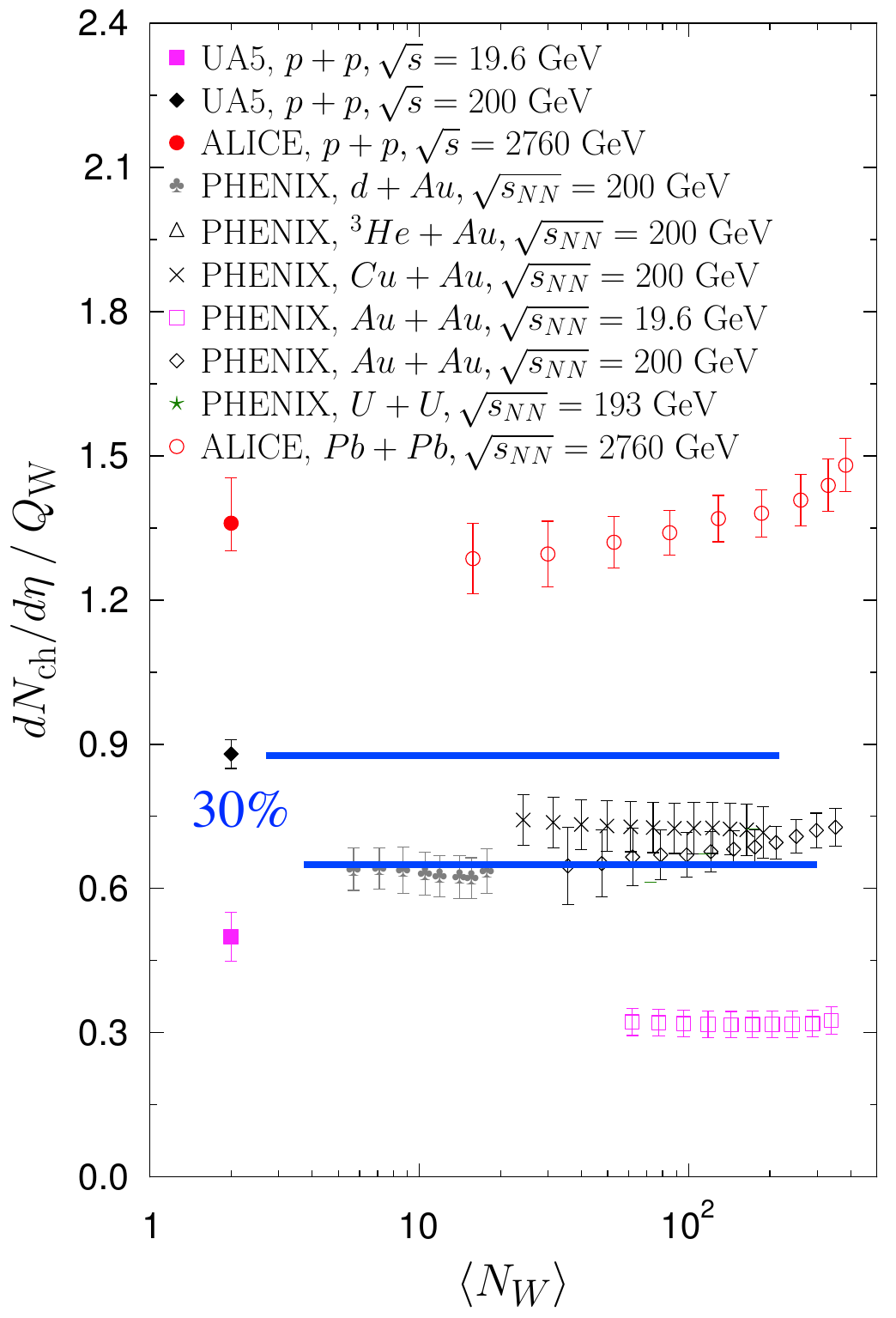}
\includegraphics[width=0.32\textwidth]{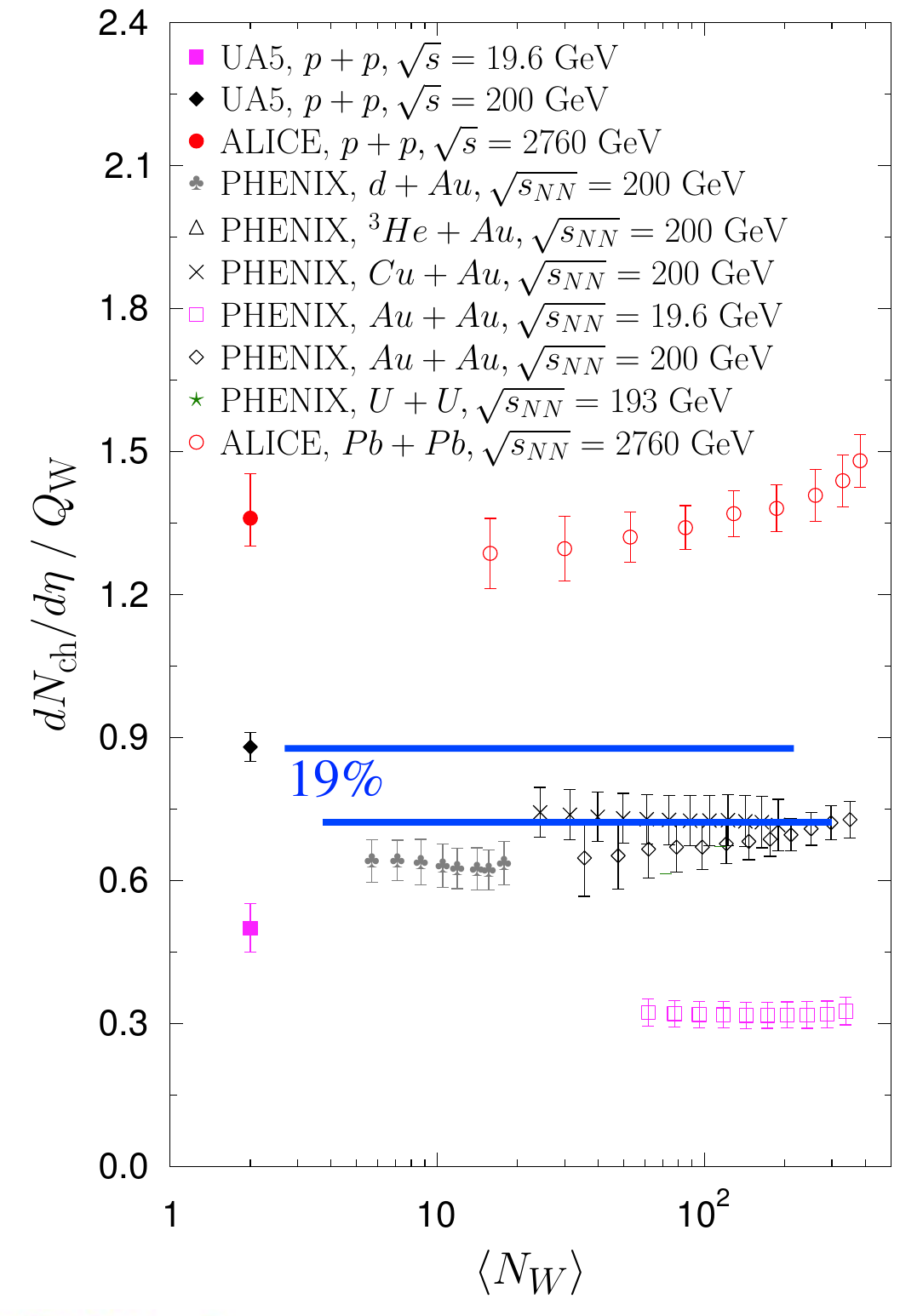}
      \vspace*{-0.5pc} 
      \caption[]{\footnotesize a) (left) Experimental multiplicity of charged hadrons per unit of pseudorapidity (at midrapidity) divided by the number of wounded quarks, $d\Nch/d\eta)/Q_W$, plotted as a function of centrality expressed via the number of wounded nucleons $N_W$~\cite{BBRPRC94}. b) (center) Same as (a) with 30\% difference from the lowest Au$+$Au measured data point to the p$+$p calculated data point indicated. c) (left) Same as (b) except 19\% difference from the top of the Au$+$Au $+1\sigma$ correlated systematic error.}\vspace*{-1.0pc}
\label{fig:Bozek}
   \end{figure}

Bozek, Broniowski and Rybczynski~\cite{BBRPRC94} did a calculation for constituent quark participants, which they call $Q_W$ (wounded quark) and for \Npart they use $N_W$ (Fig.~\ref{fig:Bozek}a). They find that the \Nqp scaling works for Alice Pb$+$Pb at \sqsn=2.76 TeV but they make the comment ``we note in Fig. 1 (Fig.~\ref{fig:Bozek}a here) that at \sqsn=200 GeV the corresponding p$+$p point is higher by about 30\% from the band of other reactions.'' However, there are several things to note in Fig.~\ref{fig:Bozek}a that Bozek {\it et al.} seem to have missed.
   \begin{figure}[!hb] 
      \centering
  \includegraphics[width=0.45\textwidth]{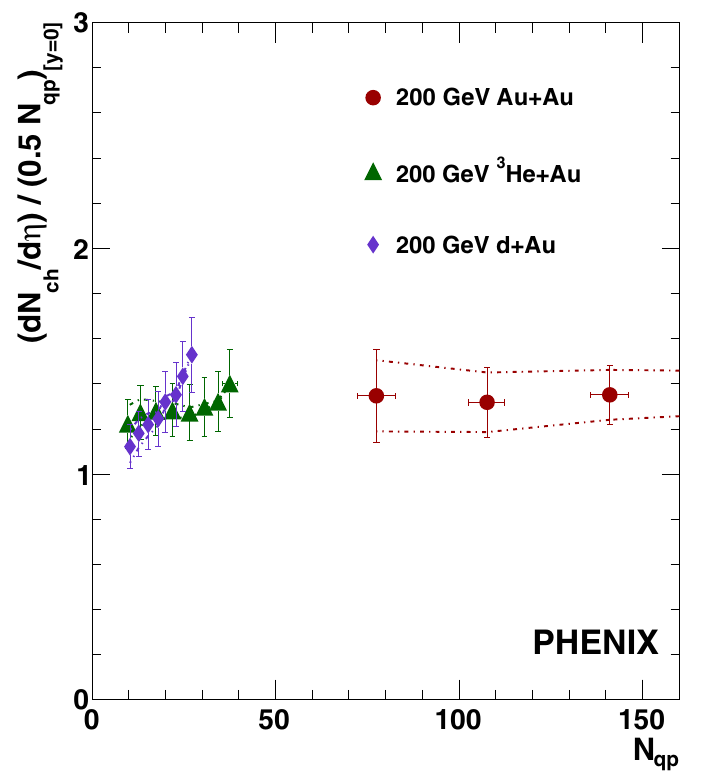}\hspace*{1.0pc}
   \includegraphics[width=0.45\textwidth]{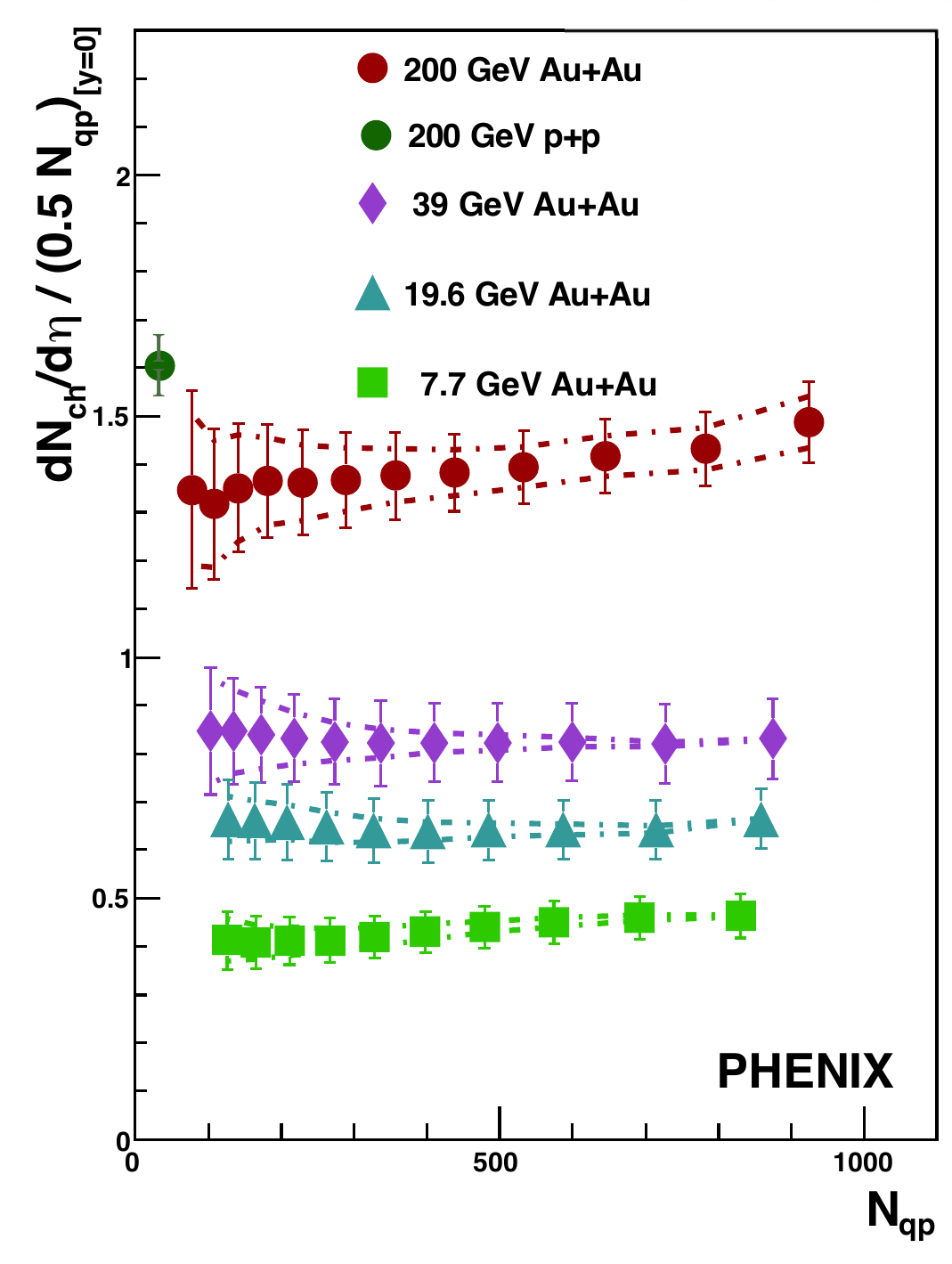}
      \vspace*{-0.5pc} 
      \caption[]{\footnotesize a)(left) $(d\Nch/d\eta)/(0.5\Nqp)$ as a function~\cite{ppg174} of \Nqp for d$+$Au, $^3$He$+$Au at \sqsn=200 GeV compared to the least central of the Au$+$Au data. b) (right) $(d\Nch/d\eta)/(0.5\Nqp)$ for \sqsn=200, 39, 19.6 and 7.7 GeV Au$+$Au data of Fig.~\ref{fig:ppg174}, with the p$+$p calculated data point added.}
\label{fig:PXwp}
   \end{figure}

Figure ~\ref{fig:Bozek}b shows that the 30\% is only valid for the lowest Au$+$Au and the d$+$Au points. The second thing to note is that they only used the tabulated data points~\cite{ppg174} which did not include the correlated systematic errors which were indicated by dashes on the figures (see the italicized sentence in the caption of Fig.~\ref{fig:ppg174} which admitedly is not as clear as it could be). If all the data points are moved up to the top of their $+1\sigma$ correlated systematic error (dashed line in Fig.~\ref{fig:ppg174}b, with a cost of $1\sigma$  (Fig.~\ref{fig:Bozek}c)), then the ratio of the p$+$p to the lowest Au$+$Au point is $1.19\pm0.17$, i.e. statistical $1.1\sigma$, plus the systematic $+1\sigma$ in quadrature, which equals a $1.5\sigma$ difference which is not significant. Regarding, the lowest d$+$Au (and $^3$He$+$Au) data points which were mentioned~\cite{BBRPRC94} in the discussion of Fig.~\ref{fig:Bozek}a, they are shown on a better scale without their correlated systematic error in Fig.~\ref{fig:PXwp}a and are in agreement with the lowest Au$+$Au data points.
\subsection{Something that we left out.}
Although the correlated systematic errors were shown on Fig.~\ref{fig:ppg174} for the Au$+$Au results, we actually did not calculate the p$+$p value of $(d\Nch/d\eta)/(0.5\Nqp)$ when we wrote the paper~\cite{ppg174}. To compare with Bozek {\it et al.}~\cite{BBRPRC94}, I used the same UA5 $\bar{\rm p}+$p $d\Nch/d\eta=2.23\pm 0.08$ at \sqs=200 GeV~\cite{UA5ZPC33} that they used together with the PHENIX2014 value~\cite{ppg174} of $\mean{\Nqp}=2.78$ for p$+$p collisions, with the result $(d\Nch/d\eta)/(0.5\Nqp)_{\rm pp}=1.60\pm0.06$ shown on Fig.~\ref{fig:PXwp}b. This is in agreement to within $\approx 1\sigma$ with all the Au$+$Au $\sqs=200$ data points in Fig.~\ref{fig:PXwp}b. It is important to note that  if the $d\Nch/d\eta$ in p$+$p collisions had been measured by PHENIX in Figs.~\ref{fig:Bozek} and \ref{fig:PXwp} then only the statistical errors $\sigma$ could be used for comparison because all the data points (including the p$+$p data) would move together by their common correlated systematic error. However, since the p$+$p $d\Nch/d\eta$ measurement is from a different experiment~\cite{UA5ZPC33} the PHENIX A$+$A measurements and correlated systematic errors are independent of the p$+$p measurement as assumed in section~\ref{sec:Bozek}.

   \begin{figure}[!hbt] 
      \centering
\raisebox{0.0pc}{\includegraphics[width=0.90\linewidth]{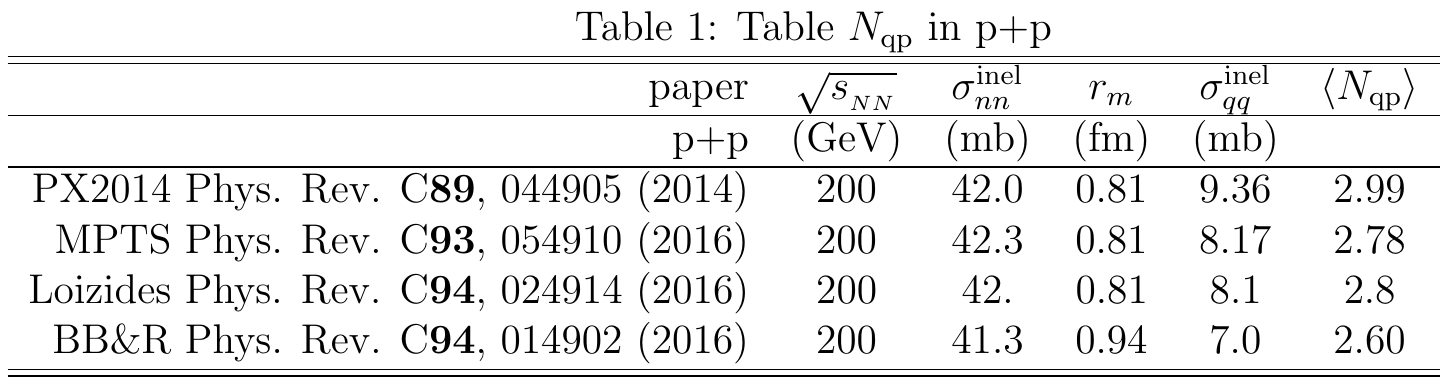}}
\raisebox{0.0pc}{\includegraphics[width=0.90\linewidth]{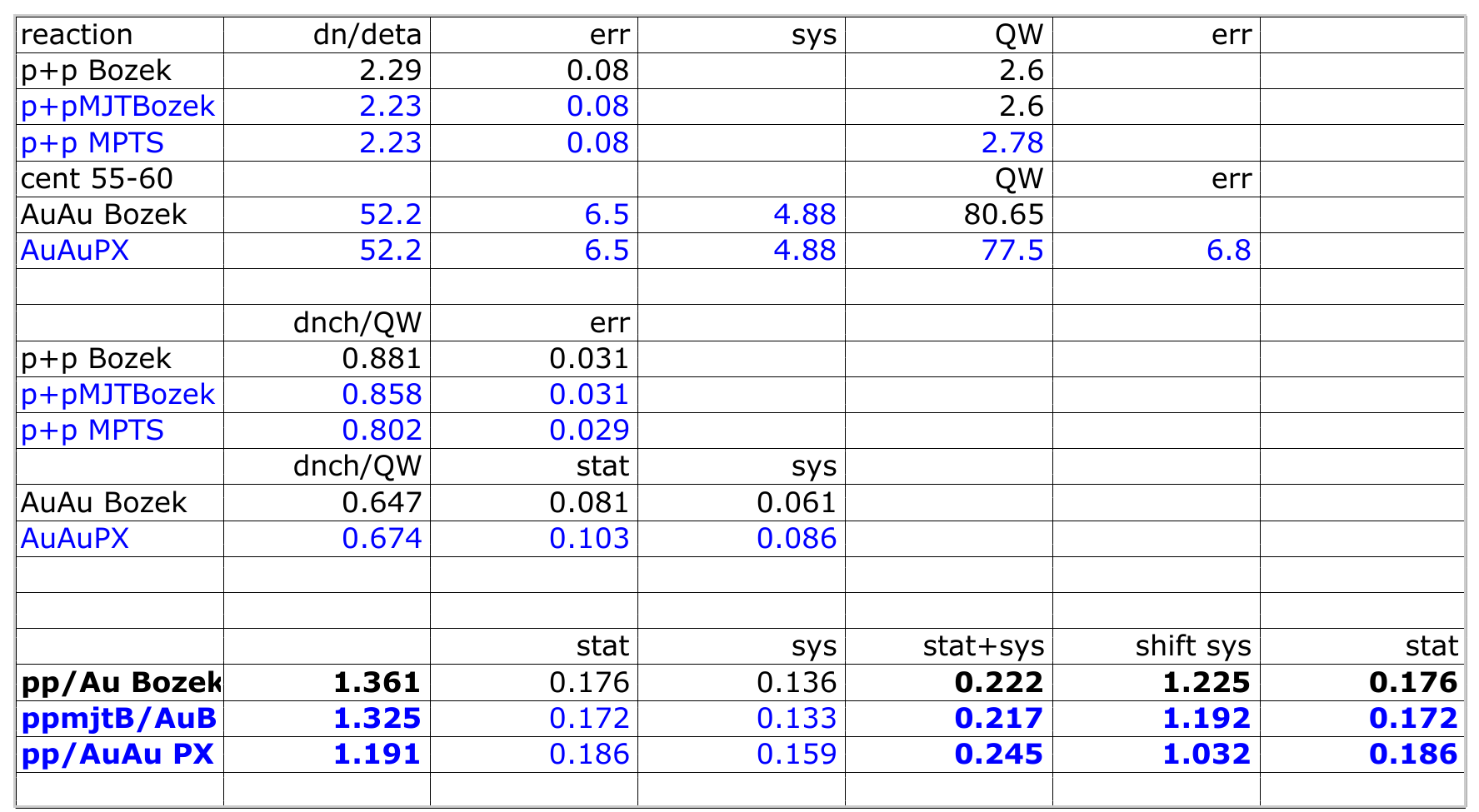}}\vspace*{-0.5pc} 
      \caption[]{\footnotesize (top) Table 1, \Nqp calculations in p$+$p collisions by four different publications indicated. (bottom) More details of the calculations in Table 1 where Bozek refers to BB\&R~\cite{BBRPRC94}, PX refers to PHENIX2014~\cite{PXPRC89} and MJTBozek refers to the p$+$p calculation in Fig.~\ref{fig:PXwp}b. }
      \label{fig:Tables}
   \end{figure} 
\subsubsection{Why are the Bozek {\it et al.}~\cite{BBRPRC94} (BB\&R) results different from the PHENIX calculation?}  
The calculation of $\mean{\Nqp}$ for p$+$p collisions seems to be the culprit. Table 1 gives values for four independent calculations in p$+$p collisions. As far as I can tell BB\&R~\cite{BBRPRC94} used $r_m=0.94$ fm for the proton rms radius in Eq.~\ref{eq:Hofstadterdipole} instead of $r_m=0.81$ fm. They also used a ``Gaussian wounding profile" for a q$+$q collision which is not the standard Glauber Monte Carlo method. The tables in Fig.~\ref{fig:Tables} give more details about the various constituent quark calculations for \sqsn=200 GeV.  
  
 \section{\large Conclusions}
\begin{itemize}
\item[i)] The constituent quark participant model (\Nqp) works at mid-rapidity for A$+$B collisions in the range ($\sim 20$ GeV) 39 GeV$\leq \sqsn\leq$ 5.02 TeV. 
\item[ii)] Experiments generally all use the same Glauber Monte Carlo method but the BB\&R~\cite{BBRPRC94} Monte Carlo is different for q$+$q scattering leading to somewhat different results.
\item[iii)] Attention must be paid to correlated systematic errors.
\item[iv)] How can the event-by-event proton radius variations and quark-quark correlations used in constituent quark Glauber calculations be measured?
\end{itemize}
\end{document}